\documentclass{PoS}
\usepackage{multirow}
\usepackage{amsmath}
\title{Parton energy loss in the reformulated weakly-coupled kinetic approach}

\ShortTitle{Parton energy loss in the reformulated weakly-coupled kinetic approach}

\author{\speaker{Tianyu Dai}\\
        Department of Physics, Duke University\\
        Durham, NC27708, USA\\
        E-mail: \email{tianyu.dai@duke.edu}}

\author{Steffen A. Bass\\
        Department of Physics, Duke University\\
        Durham, NC27708, USA\\
        E-mail: \email{bass@phy.duke.edu}}
        
\author{Jean-Fran\c cois Paquet\\
        Department of Physics, Duke University\\
        Durham, NC27708, USA\\
        E-mail: \email{jeanfrancois.paquet@duke.edu}}

\author{Derek Teaney\\
        Department of Physics \& Astronomy, Stony Brook University\\
        Stony Brook, NY11764, USA\\
        E-mail: \email{derek.teaney@stonybrook.edu}}

\abstract{Interactions between hard partons and the quark-gluon plasma range from frequent soft interactions to rare hard scatterings. The larger number of soft interactions makes possible an effective stochastic description of parton-plasma interactions in terms of drag and diffusion transport coefficients. In this work, we present a numerical implementation that builds upon this systematic division between soft and hard parton-plasma interactions.
We study the dependence of the single parton distribution on the cutoff between soft and hard parton-plasma interactions, both for small and phenomenological values of the strong coupling constant.
} %This reformulation also enables the data-driven constraints on characteristic properties of parton energy loss. }

\FullConference{International Conference on Hard and Electromagnetic Probes of High-Energy Nuclear Collisions\\
  30 September - 5 October 2018\\
  Aix-Les-Bains, Savoie, France}

\begin{document}

%%%%%%%%%%%%%%%%%%%%%%%%%%%%%%%%%%%%%%%%%%%%
%%%%%%%%%%%%%%%%%%%%%%%%%%%%%%%%%%%%%%%%%%%%
\section{Introduction \vspace{-0.3cm}}
%%%%%%%%%%%%%%%%%%%%%%%%%%%%%%%%%%%%%%%%%%%%
%%%%%%%%%%%%%%%%%%%%%%%%%%%%%%%%%%%%%%%%%%%%

The energy loss of a hard parton traveling in the quark-gluon plasma can be studied within a weakly-coupled kinetic approach.
Parton-plasma interactions is treated perturbatively, and the dynamics of well-defined quasiparticles (quarks and gluons) are described by transport equations \cite{arnold2003effective}. Both the energy gain and loss of the partons are included naturally.
% * <bass@phy.duke.edu> 2018-12-12T13:53:18.133Z:
% 
% I'm somewhat confused by this statement - our plasma evolution is calculated using hydrodynamics, which sort of implies strong interactions among the plasma constituents. It's only the hard parton - plasma constituents interaction that we assume to be weakly coupled, isn't it?
% 
% ^ <tianyudai25@gmail.com> 2018-12-12T16:48:14.917Z:
% 
% I think you are right. We should say weakly-coupled parton-medium interaction instead of weakly-coupled medium. I have changed the sentence.  
%
% ^.

Leading-order realizations of weakly-coupled effective kinetic theory, such as MARTINI~\cite{Schenke:2009gb}, generally divide parton-plasma interactions as elastic and inelastic processes. 
In Ref.~\cite{ghiglieri2016jet}, it was shown that the parton energy loss can be equivalently reformulated in terms of hard large-angle interactions and soft small-angle collisions.
A major advantage of this reformulated approach is that it could be extended to next-to-leading order~\cite{caron2009g,ghiglieri2016jet}.
There are nevertheless important benefits at leading order as well. Soft and hard parton-plasma interactions can be systematically factorized. The large number of soft interactions can be efficiently described in a stochastic approach with transport coefficients; these drag and diffusion coefficients absorb plasma effects (e.g. Debye screening) that are particularly important for soft interactions. Large-angle interactions can be treated separately with emission rates, as in previous implementations. \par
In this work, we present the first numerical implementation of this reformulated energy loss model. We study the dependence of the parton energy loss on the cutoffs dividing the soft and hard interactions. We show that the reformulated energy loss remains valid at large values of the strong coupling constant $\alpha_s$.

%%%%%%%%%%%%%%%%%%%%%%%%%%%%%%%%%%%%%%%%%%%%
%%%%%%%%%%%%%%%%%%%%%%%%%%%%%%%%%%%%%%%%%%%%
\section{
Description of the Reformulated Model \vspace{-0.3cm}}
%%%%%%%%%%%%%%%%%%%%%%%%%%%%%%%%%%%%%%%%%%%%
%%%%%%%%%%%%%%%%%%%%%%%%%%%%%%%%%%%%%%%%%%%%

The Boltzmann transport equation for a parton propagating through the quark-gluon plasma is \mbox{$p\cdot \partial f = - (p \cdot u) \mathcal{C}[f]$},
%\begin{equation}
%\frac{p\cdot \partial f}{p \cdot u} = 
%\end{equation}
where $p$ is its four-momentum, $f$ is the distribution of partons, $u$ is the velocity of the medium and $\mathcal{C}[f]$ is the collision kernel. 
In previous implementations, $\mathcal{C}[f]$ would be divided into an elastic and an inelastic term.
In the leading order reformulation~\cite{ghiglieri2016jet},
the collision kernel is divided into (i) hard elastic and inelastic interactions, (ii) diffusion, and (iii) conversion processes.

\paragraph{Hard interactions}
In the case of inelastic interactions, multiple soft interactions with the plasma induce the radiation of a parton of energy $\omega$.
% * <tianyudai25@gmail.com> 2018-12-11T04:49:26.628Z:
% 
% > In the case of inelastic interactions, multiple soft interactions with the plasma induces the radiation of a parton of energy $\omega$.
% Could we say a parton of energy $\omega$ here? $\omega$ could be small, in that case, it is not parton. 
% 
% ^ <jeanfrancois.paquet@duke.edu> 2018-12-12T20:25:35.225Z:
% 
% Technically, what you are saying is probably correct (partons cannot be "soft" kind-of by definition). In practice, it is probably an acceptable abuse of the word "parton".
%
% ^.
These induced parton emissions can be divided as large-$\omega$ and small-$\omega$ interactions by a cutoff  $\mu_\omega$ with $\mu_\omega \lesssim T$, where $T$ is the temperature of the plasma.
Small-$\omega$ inelastic interactions are absorbed into drag and diffusion coefficients. Large-$\omega$ inelastic interactions are described with emission rates, which are obtained from the AMY integral equation \cite{Arnold:2002ja}. \par

In the elastic case, a kinematic cutoff is imposed on the transverse momentum transfer $q_\perp$. The cutoff $\mu_{q_\perp}$ is chosen such that $g T \ll \mu_{q_\perp} \ll T$, with $g=\sqrt{4 \pi \alpha_s}$. 
Small-$q_\perp$ interactions are again absorbed into transport coefficients.
Large-$q_\perp$ interactions are calculated perturbatively. Because plasma effects are significant only at low $q_\perp$, it is sufficient to use vacuum matrix elements to compute the large-$q_\perp$ rate. 
Given that $p \gg T$ is an excellent approximation for phenomenological applications, we further simplify the evaluation of the large-$q_\perp$ rate by keeping only the zeroth-order term in $T/p$.
\par
\paragraph{Drag and diffusion}
Number- and identity-preserving
%\footnote{Because soft radiated particles are absorbed by the plasma, the number-preserving assumption still holds in the inelastic case.} 
soft interactions are described stochastically with drag and diffusion. Elastic as well inelastic interactions are included: because soft radiated particles are absorbed by the plasma, the number-preserving assumption still holds in the inelastic case.
% * <bass@phy.duke.edu> 2018-12-12T13:59:58.705Z:
% 
% something is missing/ambiguous in this sentence...
% 
% ^ <tianyudai25@gmail.com> 2018-12-12T16:27:45.236Z:
% 
% What we are trying to say is that we use diffusion processes to treat soft interactions. In the diffusion process, the number of the partons involved in the interaction is preserved. The identity of the parton also isn't changed in diffusion process. So, the identity non-preserving soft interactions should be treated by conversion process. 
%
% ^ <tianyudai25@gmail.com> 2018-12-12T16:35:09.010Z:
% 
% Could I change it to: Soft interactions from both elastic and inelastic part are treated with a diffusion process. In the diffusion process, the number and the identity of the partons involved in the interactions is preserved. The identity non-preserving soft interactions are included in conversion processes. 
%
% ^ <jeanfrancois.paquet@duke.edu> 2018-12-12T20:47:56.853Z:
% 
% I changed it for something along what you wrote:
% "Number- and identity-preserving soft interactions are described stochastically with drag and diffusion. Elastic as well inelastic interactions are included: because soft radiated particles are absorbed by the plasma, the number-preserving assumption still holds in the inelastic case."
%
% ^.
The diffusion processes can be described by a Fokker-Planck equation: 
\begin{equation}
\label{FP}
\mathcal{C}^{diff}[f] = -\frac{\partial}{\partial p^i}\left[\eta_D(p)p^if(\vec{p})\right]
-\frac{1}{2}\frac{\partial^2}{\partial p^i\partial p^j}\left\{\left[\hat{p}^i\hat{p}^j \hat{q}_L(p)+\frac{1}{2}\left(\delta^{ij}-\hat{p}^i\hat{p}^j\right)\hat{q}(p)\right]f(\vec{p})\right\}
\end{equation}
% \begin{equation}
% \label{FP}
% \begin{split}
% \mathcal{C}^{diff}[f] = &-\frac{\partial}{\partial p^i}\left[\eta_D(p)p^if(\vec{p})\right]\\
% &-\frac{1}{2}\frac{\partial^2}{\partial p^i\partial p^j}\left\{\left[\hat{p}^i\hat{p}^j\left(\hat{q}^{inel}_L(p)+\hat{q}^{elas}_L(p)\right)+\frac{1}{2}\left(\delta^{ij}-\hat{p}^i\hat{p}^j\right)\hat{q}(p)\right]f(\vec{p})\right\}
% \end{split}
% \end{equation}
where $\eta_D$ is the drag, and $\hat{q}_L(p)$ and $\hat{q}$ are the longitudinal and transverse momentum diffusion coefficients. 
%$\hat{q}_L(p)=\hat{q}^{inel}_L(p)+\hat{q}^{elas}_L(p)$, where $\hat{q}^{inel}_L(p)$ is attributed to inelastic interactions and $\hat{q}^{elas}_L(p)$ is attributed to elastic interactions. 
\par
The latter are calculated perturbatively while the drag coefficient $\eta_D$ is obtained by the Einstein relation. 
The elastic contribution to both $\hat{q}$ and $\hat{q}_L$ can be found in Ref.~\cite{ghiglieri2016jet}.
%Both $\hat{q}$ and $\hat{q}_L^{elas}$ are $\mathcal{O}(g^2)$; their value can be found in Ref.~\cite{ghiglieri2016jet}.
%\begin{equation}
%\hat{q} = \frac{g^2C_RTm_D^2}{4\pi}\ln\left[1+(\frac{\mu_{q_\perp}}{m_D})^2\right]
%\end{equation}
%where $m_D$ is the leading order Debye mass. The longitudinal momentum broadening contributed by elastic energy loss is: 
%\begin{equation}
%\hat{q}^{elas}_L = \frac{g^2C_RTM_\infty^2}{4\pi}\ln\left[1+(\frac{\mu_{\tilde{q}_\perp}}{M_\infty})^2\right]
%\end{equation}
%where $M_\infty$ is the gluon asymptotic thermal mass. 
%\begin{equation}
%\hat{q}^{inel}_L = \frac{(2-\ln2)g^4C_RC_AT^2\mu_\omega}{4\pi^3}
%\end{equation}\par
%\begin{equation}
%\eta_D(p) = \frac{\hat{q}_L}{2Tp}+\frac{1}{2p^2}(\hat{q}-2\hat{q}_L)
%\end{equation}
Inelastic interactions also contribute to the longitudinal diffusion: $\hat{q}^{inel}_L = \frac{(2-\ln2)g^4C_RC_AT^2\mu_\omega}{4\pi^3}$. \par

 \par
%\begin{equation}
%\Delta \vec{x} = \frac{\vec{p}}{E_{\vec{p}}}\Delta t
%\end{equation}
%\begin{equation}
%\Delta \vec{p} = -\eta_D(\vec{p})\vec{p}\Delta t + \vec{\xi}(t)\Delta t
%\end{equation}
%where $\xi$ is a thermal random force: 
%\begin{equation}
%\left<\xi_i(t)\xi_j(0)\right>=(\hat{q}_L-\frac{1}{2}\hat{q})\hat{p}_i\hat{p}_j%+\frac{1}{2}\hat{q}\delta_{ij})\delta(t)
%\end{equation}
\paragraph{Conversion}
An additional type of interactions are conversions, in which the incoming parton changes identity.
%They can be thought of a small-angle interactions in which identify is not preserved. 
The leading contribution to this process, both in the elastic and inelastic cases, is suppressed by $\mathcal{O}(T/p)$ compared to the large-angle and diffusion processes. Consequently, their contribution is not included in this work. 
%We did verify that their contribution was negligible.

%When the transverse momentum transfer $q_\perp$ of the conversion process is larger than the cutoff $\mu_{q_\perp}$, the contribution of conversion process to QCD matrix elements is $O(T/p)$, which is negligible. When $q_\perp$ is smaller than $\mu_{q_\perp}$, the identities of the partons are changed according to the conversion rates listed in \cite{ghiglieri2016jet}, which is also $\mathcal{O}(T/p)$. Conversion rate of the inelastic conversion processes are also suppressed by $\mathcal{O}(T/p)$ \cite{ghiglieri2016jet}. 

\paragraph{Reformulation}
The reformulation can be written as: 
\begin{equation}
\mathcal{C} = \mathcal{C}^{2\leftrightarrow 2} + \mathcal{C}^{1\leftrightarrow 2} = \mathcal{C}^{large-\omega}\left(\mu_\omega\right)+\mathcal{C}^{large-q_\perp}\left(\mu_{q_\perp}\right)+\mathcal{C}^{diff}\left(\mu_\omega, \mu_{q_\perp}\right)+\mathcal{C}^{conv}\left(\mu_{q_\perp}\right)
\label{reformulation}
\end{equation}
%Treatments of different processes in the reformulated model are summarized in the Table \ref{tab}. 
Every term has a cutoff dependence, which cancels out when added together. In what follows we test this cutoff independence numerically, for the first time.

%%%%%%%%%%%%%%%%%%%%%%%%%%%%%%%%%%%%%%%%%%%%
% \begin{table}
% \caption{Treatments of reformulated energy loss}
%   \centering
%   \begin{tabular}{ | c | c | c | c | }
%     \hline
%      \multicolumn{2}{| c |}{reformulated interactions} & identity preserving & identity non-preserving \\ 
%      \hline
%     \multirow{2}{*}{soft interactions} & small-$q_\perp$ elastic & \multirow{2}{*}{Diffusion processes} & \multirow{3}{*}{$\frac{T}{p}$ suppressed} \\
%     \cline{2-2}
%     & small-$\omega$ inelastic & & \\
%     \cline{1-3}
%     \multirow{2}{*}{hard interactions} & large-$q_\perp$ elastic & vacuum matrix elements &  \\
%     \cline{2-4}
%     & large-$\omega$ inelastic & \multicolumn{2}{| c |}{AMY integral equations}\\
%     \hline
%   \end{tabular}
%   \label{tab}
% \end{table}
%%%%%%%%%%%%%%%%%%%%%%%%%%%%%%%%%%%%%%%%%%%%

%\section{Cutoff Independence Test in a Static Medium}
%%%%%%%%%%%%%%%%%%%%%%%%%%%%%%%%%%%%%%%%%%%%
%%%%%%%%%%%%%%%%%%%%%%%%%%%%%%%%%%%%%%%%%%%%
%\section{Testing the reformulated energy loss in a static medium}
\section{Numerical implementation \& results in a static medium \vspace{-0.3cm}}
%%%%%%%%%%%%%%%%%%%%%%%%%%%%%%%%%%%%%%%%%%%%
%%%%%%%%%%%%%%%%%%%%%%%%%%%%%%%%%%%%%%%%%%%%

To implement the reformulated energy loss model described in the previous section, we used the public version of the JETSCAPE framework~\cite{kauder2018jetscape}, building upon the existing implementation of MARTINI~\cite{Schenke:2009gb} already in the framework.
%in the JETSCAPE framework was straightforward to modify to handle the inelastic large-$\omega$ and the elastic large-$q_\perp$ processes. 
The Fokker-Planck equation describing the soft interactions (\ref{FP}) was added as a Langevin equation, solved with the pre-point Ito scheme. 
%This allows for a 
%which is a powerful and convenient tool for jet energy loss analysis. The reformulated model is added as a separate external module in a modified version of the public JETSCAPE code.
\par

Strictly speaking, the reformulation of parton energy loss Eq. (~\ref{reformulation}) is valid in the small coupling limit. Independence on the cutoffs separating soft and hard interactions ($\mu_{q_\perp}$ and $\mu_\omega$) is not assured in phenomenological applications, where a large coupling is used. We discuss this cutoff dependence in what follows.  Because the cancellation of the cutoff is essentially independent in the elastic and inelastic cases, we look at them separately.

\paragraph{Inelastic energy loss}

\begin{figure}[!h]
 \centering
 \includegraphics[width=\linewidth]{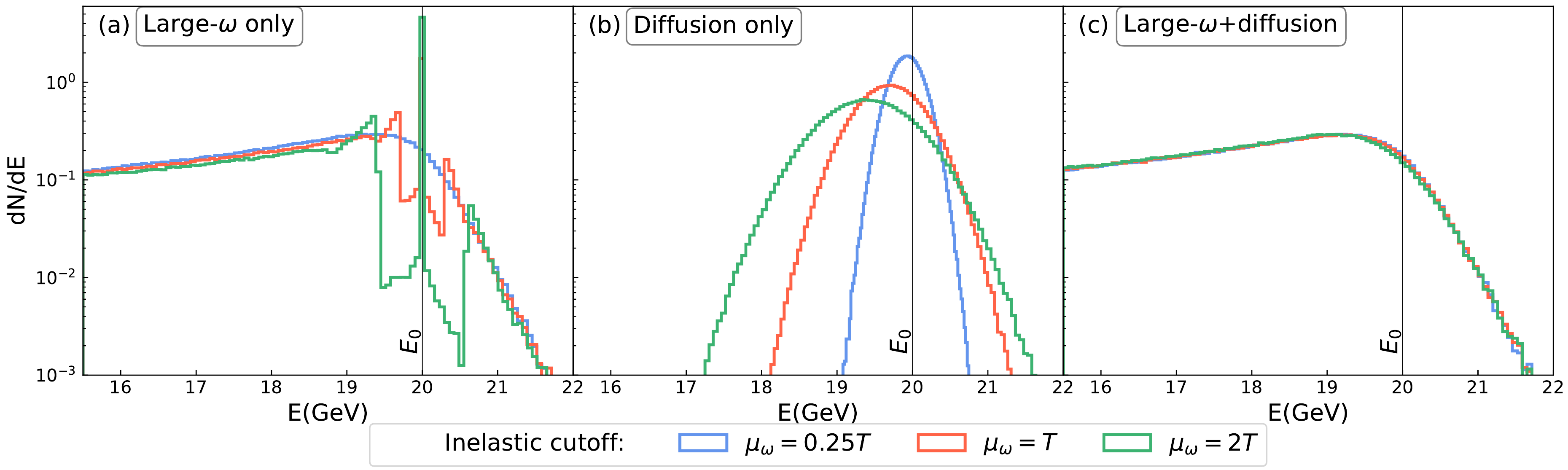}
 \caption{Energy distribution of a 20 GeV gluon losing energy through inelastic processes only in a 1 fm/c static quark-gluon plasma at T = 300 MeV, for different $\mu_\omega$. For  (a) large-$\omega$ only, (b) small-$\omega$ only, and (c) combined energy loss. }
 \label{fig_inel}
\end{figure}

We set $\alpha_s = 0.3$  and simulate the propagation of a 20 GeV gluon traveling for 1 fm/c in an infinite static medium. Only inelastic energy loss is included. The temperature of the plasma is 300 MeV. We vary the inelastic cutoff $\mu_\omega$ to the value $\mu_\omega = 0.25T, \, T, \, 2T$. 
The leading-gluon energy distribution  
is shown in Figure \ref{fig_inel}, first separately for the large-$\omega$ interactions and the drag and diffusion, and then combined.

%[The cutoff should satisfy the condition that interactions with $\omega < \mu_\omega$ are frequent enough, in the sense that diffusion process is still a good description.] 

In the large-$\omega$ case (Figure \ref{fig_inel}(a)), all soft radiations below the cutoff $\mu_\omega$ are forbidden; inevitably, the energy distribution around the initial parton energy is found to depend on $\mu_\omega$. In the drag and diffusion case (Figure \ref{fig_inel}(b)), because the transport coefficients increase linearly with $\mu_\omega$, the parton energy loss also increases on $\mu_\omega$.
Once combined, Figure \ref{fig_inel}(c), the $\mu_\omega$ dependence of the parton energy loss cancels out.

In fact, when $\mu_\omega$ is decreased, we reach the limit where the effect of small-$\omega$ radiation is negligible. This is the limit in which inelastic energy loss is typically implemented. The reformulated energy loss allows for this cutoff to be increased and varied.

We verified that results similar to Figure \ref{fig_inel} were obtained with (i) a smaller coupling constant, (ii) different initial parton energies, (iii) a quark propagating instead of a gluon, and (iv) a realistic hydrodynamic medium used instead of a brick.

\begin{figure}[!tbh]
 \centering
 \includegraphics[width=\linewidth]{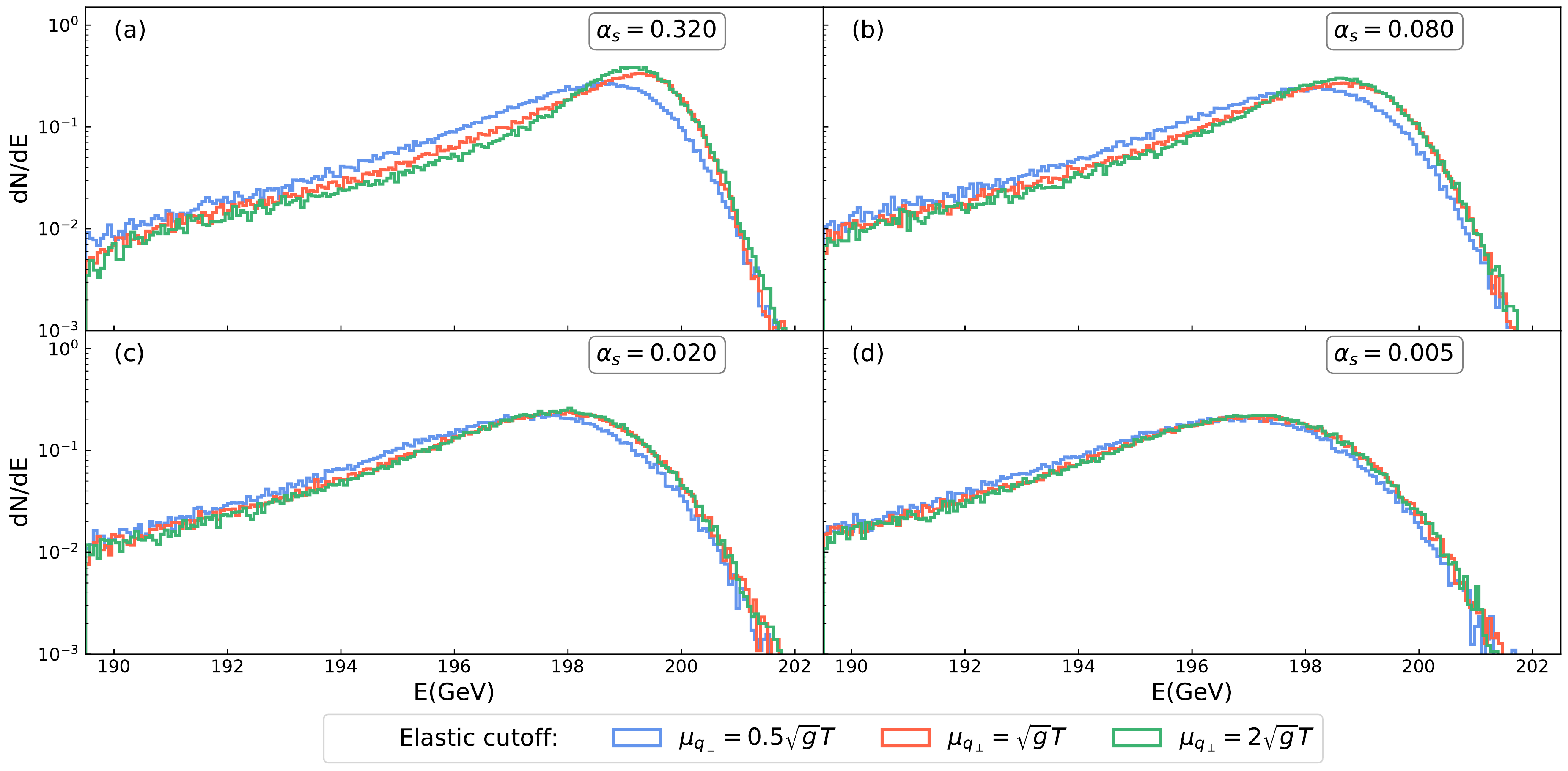}
 \caption{Energy distribution of a gluon with initial energy $E_0 = 200$ GeV losing energy through elastic processes in a static plasma with $T = 300$ MeV, for a propagation length $\tau=\alpha_s^{-2} 0.3^2$~fm/c at different elastic cutoff $\mu_{q_\perp}$. The results are shown for four different coupling constants $\alpha_s$.}
 \label{fig_elas}
\end{figure}

\paragraph{Elastic energy loss}

Figure \ref{fig_elas}(a) shows the energy distribution of a 200 GeV gluon propagating for 1 fm/c in the same medium as used in the previous example: infinite length, $T = 300$~MeV. Only elastic energy loss is included. We show only the combined result with both large-angle scattering and drag and diffusion. At large coupling, the dependence on the elastic cutoff $\mu_{q_\perp}$ is modest, although larger than found in the inelastic case. This larger cutoff dependence, in the large coupling limit, is likely a consequence of our use of vacuum matrix elements in the large $q_\perp$ elastic energy loss calculation, instead of screened matrix elements. In the small coupling regime, screening effects are small for large-$q_\perp$ interactions, and the use of vacuum matrix elements is enough. As expected, we find that the cutoff independence is recovered when the coupling is reduced\footnote{We fix the propagation length $\tau = \alpha_s^{-2} 0.3^2$~fm/c to make the number of elastic collisions the same for tests with different coupling constants.}, as shown in Figure \ref{fig_elas}(b-c). 
%We find however that the cutoff independence is recovered, when the coupling is reduced, as expected\footnote{We fix $\alpha_s^2\tau = 0.3^2$ to make the number of elastic collisions the same for tests with different coupling constants.}, as shown in Figure \ref{fig_elas}(b-c).
%%%%%%%%%%%%%%%%%%%%%%%%%%%%%%%%%%%%%%%%%%%%
%%%%%%%%%%%%%%%%%%%%%%%%%%%%%%%%%%%%%%%%%%%%
%\section{Summary}
\section{Summary \& outlook \vspace{-0.3em}}
%%%%%%%%%%%%%%%%%%%%%%%%%%%%%%%%%%%%%%%%%%%%
%%%%%%%%%%%%%%%%%%%%%%%%%%%%%%%%%%%%%%%%%%%%
The reformulated parton energy loss from Ref.~\cite{ghiglieri2016jet} provides a systematic factorization of soft and hard interactions in the weakly-coupled limit. The numerical implementation presented in this work indicates that this factorization still holds well at large coupling and can be used in phenomenological studies.

Naturally one benefit of this reformulation is the possibility of extending it to next-to-leading order~\cite{ghiglieri2016jet}. A separate benefit is phenomenological: this methodical separation of the transport (soft) sector of the parton energy loss from hard interactions provides a framework for data-driven constraints on the drag and diffusion coefficient of partons. We expect both of these directions to have important applications for the study of jets in heavy ion collisions.
%We implemented the reformulated parton energy loss model and compute the energy loss in a static medium with different coupling constants and different cutoff for both gluon and quark. The energy loss of both elastic and inelastic interactions is independent on the cutoff between soft interactions and hard interactions. \par
%We expect the reformulated model to be more efficient and flexible. Drag and diffusion processes are factorized out, which enables us to extend the model from perturbative regime to non-perturbative regime. The parameterizable nature of drag and diffusion coefficients also makes it convenient for us to apply data-driven constraints. This reformulated approach can also be extended to the next-to-leading order. 
%\acknowledgments
\vspace{-0.3em}
\paragraph{Acknowledgements}
We are grateful to Jacopo Ghiglieri, Weiyao Ke and Yingru Xu for their help with this project. We thank Bjoern Schenke and Heikki M\"antysaari for their collaboration in the early stage of this project, and the JETSCAPE Collaboration for their assistance with the JETSCAPE framework.
This work was supported by the National Science Foundation under Award Number ACI-1550225 (T.D.), the U.S. Department of Energy  under Award Numbers DE-FG02-05ER41367 (S.A.B., T.D., J.-F.P.) and DE-FG02-88ER40388 (J.-F.P., D.T.). 

\bibliographystyle{JHEP}
\bibliography{HP18ProBib}
%\begin{thebibliography}{99}
%\bibitem{arnold2003effective}
%....

%\end{thebibliography}

\end{document}